\documentclass[12pt,a4paper,twoside]{article}
\def\runninghead#1#2{\pagestyle{myheadings}
\markboth{{\protect\footnotesize\it{\quad #1}}\hfill}
{\hfill{\protect\footnotesize\it{#2\quad}}}} \headsep=15pt
\usepackage{amssymb}
\usepackage{latexsym}
\newcommand{\be}{\begin{equation}}
\newcommand{\ee}{\end{equation}}
\newcommand{\bea}{\begin{eqnarray}}
\newcommand{\eea}{\end{eqnarray}} %
 \title{Torsion and Axial  Current}
 \author{Prasanta Mahato\footnote{Narasinha Dutt College,
         Howrah, West Bengal, India 711 101}\hspace{2mm}\footnote{E-mail: pmahato@dataone.in}}
\date{}
\begin{document}
\runninghead{\underline{Prasanta Mahato}}{\underline{Torsion and Axial  Current}}
%\begin{document}
%\author{Prasanta Mahato}
%\email[email: ]{ pmahato@vsnl.net}
%\affiliation{ Department of Mathematics, Narasinha Dutt College\\         Howrah, West Bengal, India 711 101         }
%\address{ Department of Mathematics, Narasinha Dutt College\\
        % Howrah, West Bengal, India 711 101
         %}
 %\date{\today}
\maketitle
\setcounter{footnote}{0}
%\onehalfspacing
%\singlespacing
 \begin{abstract}
 The role of  torsion and a scalar field $\phi$
   in gravitation, especially, in the presence of a Dirac field  in the background of a particular class of the Riemann-Cartan geometry is considered here.
 Recently,  a Lagrangian density with Lagrange multipliers has been proposed by the author which has been obtained  by picking some particular terms from the $SO(4,1)$ Pontryagin  density,  where the scalar field $\phi$ causes  the de Sitter connection to have the proper dimension of a gauge field. In this formalism, conserved axial vector matter current can be constructed, irrespective of any gauge choice,  in any manifold having arbitrary background geometry. This current is not a Noether current.

\vspace{2mm}
   \noindent \textbf{\uppercase{pacs:}} 04.20.Fy, 04.20.Cv, 11.40.-q
   
\vspace{2mm}
   \noindent \textbf{\uppercase{key words:}}   Nieh-Yan Density, Torsion, Axial Current, Anomaly

 \end{abstract}
 %\maketitle
 %\eject
%\doublespacing
%\triplespacing
%\tableofcontents
%*************************************************************************************
\section{Introduction}

While we would like to believe that the fundamental laws of Nature are symmetric, a completely symmetric world would be dull, and as a matter of fact, the real world is not perfectly symmetric. More precisely, we want the Lagrangian, but not the world described by the Lagrangian, to be symmetric. Indeed, a central theme of modern physics is the study of how symmetries of the Lagrangian are broken. 

  It is a remarkable result of differential geometry that
 certain global features of a manifold are determined by some local invariant densities.
These topological invariants have  an important property in common -  they are total divergences
and in any local theory  these invariants,  when treated as Lagrangian densities, contribute
 nothing to the Euler-Lagrange equations.  Hence in a local theory only few parts, not the
  whole part, of these invariants can be kept in a Lagrangian density. Sometimes ago, in
this direction, a gravitational Lagrangian was
proposed\cite{Mah02a}, where a
 Lorentz invariant part of the de Sitter Pontryagin density, i.e. having broken de Sitter symmetry, was treated as
 the Einstein-Hilbert Lagrangian.  By this way the role of torsion in the underlying manifold
has become multiplicative
   rather than additive one and  the  Lagrangian  looks like
    $torsion \otimes curvature$.  In other words - the additive torsion is decoupled from the
   theory but not the multiplicative one. This indicates that torsion is uniformly nonzero
   everywhere. In the geometrical sense, this implies that
   micro local space-time is such that at every point there is a
   direction vector (vortex line) attached to it. This effectively
   corresponds to the non commutative geometry having the manifold
   $M_{4}\times Z_{2}$, where the discrete space $Z_{2}$ is just
   not the two point space\cite{Con94} but appears as an attached direction vector. This has direct relevance in the quantization of a fermion where the discrete space appears as the internal space of a particle\cite{Gho00}. Considering torsion and torsion-less
  connection as independent fields\cite{Mah04}, it has been found that $\kappa$  of Einstein-Hilbert Lagrangian, appears as an integration constant in such a way that it has been found to be linked with the topological Nieh-Yan density of $U_{4}$ space.
  If we consider axial vector torsion together with a scalar field $\phi$ connected to a local  scale factor\cite{Mah05,Mah07a}, then the Euler-Lagrange equations not only give the constancy of the gravitational constant but they also link, in laboratory scale, the mass of the scalar field with the Nieh-Yan density and, in cosmic scale of FRW-cosmology, they predict only three kinds of the phenomenological energy density representing mass, radiation and cosmological constant.
Recently it has been shown that\cite{Mah07b}, using field equations of all fields except the frame field, the starting Lagrangian  reduces to a generic $f(\mathcal{R})$ gravity Lagrangian  which, for FRW metric,  gives standard FRW cosmology. But for non-FRW metric, in particular of Ref.\cite{Lob07a}, with some particular choice of the functions of the scalar field $\phi$ one gets $f(\mathcal{R})=f_0\mathcal{R}^{1+v^2_{tg}}$, where $v_{tg}$ is the constant tangential velocity of the stars and gas clouds in circular orbits in the outskirts of spiral galaxies. With this choice of functions of $\phi$ no dark matter is required to explain flat galactic  rotation curves

We know that, an anomaly is called
a phenomenon in which the given symmetry and the corresponding conservation law of a classical filed-theory Lagrangian are violated as we pass
to quantum theory. The reason for such violation lies in the singularity of quantum field
operators at small distances, such that finding the physical quantities requires fixing not only
the Lagrangian but also the renormalization procedure\cite{Fuj04}. 
It is also well known that the existence of anomalies can be
attributed to the topological properties of the background
where the quantum system is defined\cite{Egu80}. In particular, for
a massless spin one-half field in an external  gauge field \textbf{G}, the anomaly for the conservation
law of the chiral current is proportional to the
Pontryagin form for the gauge group,
\begin{eqnarray}
	<d J_5>=\frac{1}{4\pi^2}Tr(F\wedge F)\label{eqn:01}
\end{eqnarray}
Kimura \cite{Kim69}, Delbourgo and Salam \cite{Sal72}, and Eguchi
and Freund \cite{Fre76} evaluated the quantum violation of the
chiral current conservation in a four dimensional Riemannian
background without torsion, finding it proportional
to the Pontryagin density of the manifold,
\begin{eqnarray}
	<d J_5>=\frac{1}{8\pi^2}R^{ab}\wedge R_{ab}
\end{eqnarray}This result was also supported by the computation of
Alvarez-Gaum\'e and Witten \cite{Wit84} and Bonora, Pasti \&
Tonin \cite{Ton86}.

The question then naturally arises as to whether the
torsional invariants can produce similar physically observable
effects \cite{Zan91}.
In a space time with non vanishing torsion there can occur
topologically stable configurations associated with the frame bundle
which are independent of the curvature. The relevant topological
invariants are integrals of local scalar densities first discussed by
Nieh and Yan. In four dimensions, the Nieh-Yan form $N=(T^a \mbox{\tiny $\wedge$} T_a - R_{ab}\mbox{\tiny $\wedge$} e^a\mbox{\tiny $\wedge$} e^b)$ is the only closed 4-form invariant under local Lorentz
rotations associated with the torsion of the manifold.  The chiral anomaly in a
four-dimensional space time with torsion has been shown to contain a
contribution proportional to $N$, besides the usual Pontryagin
density related to the space time curvature\cite{Cha97}. 

It is always possible to define an axial current three-form from a Chern-Simons class current  for a massless spin one-half field in an external  gauge field G\cite{Wei96,Kak93}, given by 
\begin{eqnarray}
	\widetilde{J}_5=J_5-\frac{1}{4\pi^2}(A\wedge F-\frac{1}{3}A\wedge A\wedge A).
\end{eqnarray} Using equation (\ref{eqn:01}) we see that the current $\widetilde{J}_5$ is conserved  but  not gauge invariant. In standard model, anomalies from leptonic sector cancel anomalies from quark sector. Questions naturally arises, whether it is possible to construct an axial current in curved space-time with torsion, irrespective of any particular model of internal symmetry group,  which is conserved and as well as gauge invariant. This article has been prepared to study this possibility of obtaining a conserved axial current following the results obtained in Ref.\cite{Mah07a}. 

In section-2 and section-3 we briefly describe the geometrical approach of the  formalism of Ref.\cite{Mah07a}. In section-4 we also give a short desfription of results obtained in Ref.\cite{Mah07a}. Section-4 is devoted to new results. Last section is kept for discussion.

%%%%%%%%%%%%%%%%%%%%%%%%%%%%%%%%%%%%%%%%%%%%%%%%%%%%%%%%%%%%%%%%%%%%%%%%%%%%%%%%

 \section{Axial Vector Torsion and Gravity}

Cartan's structural equations for a Riemann-Cartan space-time $U_{4}$ are given by
\cite{Car22,Car24}
 \begin{eqnarray}T^{a}&=& de^{a}+\omega^{a}{}_{b}\wedge e^{b}\label{eqn:ab}\\
 R^{a}{}_{b}&=&d\omega^{a}{}_{b}+\omega^{a}{}_{c}\wedge \omega^{c}{}_{b},\label{eqn:ac}
\end{eqnarray}
here $\omega^{a}{}_{b}$ and e$^{a}$ represent the spin connection
and the local frame respectively.

In $U_{4}$ there exists  two invariant closed four forms. One is the well
known Pontryagin\cite{Che74,Che71} density \textit{P} and the
other is the less known Nieh-Yan\cite{Nie82} density \textit{N}
given by
\begin{eqnarray} \textit{P}&=& R^{ab}\wedge R_{ab}\label{eqn:ad}\\  \mbox{and} \hspace{2 mm}
 \textit{N}&=& d(e_{a}\wedge T^{a})\nonumber\\
&=&T^{a}\wedge T_{a}- R_{ab}\wedge e^{a}\wedge
e^{b}.\label{eqn:af}\end{eqnarray}

Here we consider a particular class of the Riemann-Cartan geometry where only the axial vector part of the torsion is nontrivial.
Then, from  (\ref{eqn:af}),  one naturally gets the Nieh-Yan
density 
\begin{eqnarray} N&=&-R_{ab}\wedge e^{a}\wedge e^{b}=-{}^* N\eta\hspace{2 mm},
\label{eqn:xaa}\\
 \mbox{where} \hspace{2 mm}\eta&:=&\frac{1}{4!}\epsilon_{abcd}e^{a}\wedge e^b\wedge
e^c\wedge e^d\end{eqnarray}is the invariant volume element.  It follows that    ${}^*N$,
the Hodge dual of $N$, is a scalar density of dimension $(length)^{-2}$.

We can combine the spin connection and the vierbeins multiplied by a scalar field together in a connection for $SO(4,1)$, in the tangent space, in the
form
\begin{eqnarray}W^{AB}&=&\left
[\begin{array}{cc}\omega^{ab}&\phi e^{a}\\- \phi e^{b}&0\end{array}\right],\label{eqn:aab}
\end{eqnarray}
where $a,b = 1,2,..4$; $A,B = 1,2,..5$ and $\phi$ is a variable
parameter of dimension $(length)^{-1}$ and Weyl weight $(-1)$, such
that, $\phi e^a$ has the correct  dimension and conformal weight
of the de Sitter boost part of the $SO(4,1)$ gauge connection. In
some earlier works\cite{Cha97,Mah02a,Mah04}  $\phi$ has been
treated as an inverse length constant. In a recent
paper\cite{Mah05} $\phi$ has been associated, either in laboratory
scale or in cosmic sale, with a local energy scale. In laboratory
scale its coupling with torsion gives the mass term of the scalar
field and in cosmic scale it exactly produces the phenomenological
energy densities of the FRW universe.  The gravitational Lagrangian, in this approach, has been defined to be
\begin{eqnarray}\mbox{$\mathcal{L}$}_{G}&=& -\frac{1}{6}({}^*N \mbox{$\mathcal{R}$}\eta  +\beta \phi^2 N)+{}^*(b_a\wedge
\bar{\nabla} e^{a})(b_a\wedge
\bar{\nabla} e^{a}) \nonumber\\&{}&- f(\phi)d\phi\wedge{}^*d\phi -h(\phi)
\eta,\label{eqn:abcd1}\end{eqnarray} where
  *   is Hodge duality operator,    $\mathcal {R}$$\eta=\frac{1}{2}\bar{R}^{ab}\wedge\eta_{ab}$, $\bar{R}^b{}_a=d\bar{\omega}^b{}_a+\bar{\omega}^b{}_c\wedge \bar{\omega}^c{}_a$, $\bar{\omega}^{a}{}_{b}=\omega^{a}{}_{b}-T^{a}{}_{b}$, $ T^a=e^{a\mu}T_{\mu\nu\alpha}dx^\nu\wedge dx^\alpha$, $T^{ab}=e^{a\mu}e^{b\nu}T_{\mu\nu\alpha}  dx^\alpha$, $T=\frac{1}{3!}T_{\mu\nu\alpha}dx^\mu\wedge dx^\nu\wedge dx^\alpha$, $N=6dT$, $\eta_a=\frac{1}{3!}\epsilon_{abcd}e^b\wedge e^c\wedge e^d$ and $\eta_{ab}={}^*(e_a\wedge e_b)$. Here $\beta$ is a dimensionless coupling constant, $\bar{\nabla}$
represents covariant differentiation with respect to the connection one form $\bar{\omega}^{ab}$, $b_{  a}$ is a two form with
one internal index and of dimension $(length)^{-1}$ and $f(\phi)$,  $h(\phi)$ are unknown functions of $\phi$ whose
forms are to be determined subject to the geometric structure of the manifold.
  The geometrical implication of the first term, i.e. the $torsion \otimes curvature$\footnote{An important advantage of this part of the Lagrangian is that - it is a
   quadratic one with respect to the field derivatives and this
   could be valuable in relation to the quantization program of gravity like other gauge theories of
   QFT.} term, in the Lagrangian $\mathcal{L}\mbox{$_{G}$}$   has already been  discussed in the beginning.

  The Lagrangian $\mathcal{L}\mbox{$_{G}$}$
     is  only Lorentz invariant  under rotation in the tangent space where  de Sitter
     boosts are not permitted. As a consequence $T$ can be treated independently of $e^a$
     and $\bar{\omega}^{ab}$.
Here we note that, though torsion one form $T^{ab}=\omega^{ab}-\bar{\omega}^{ab}$ is a part of
the $SO(3,1)$ connection, it
 does not transform like a connection  form under $SO(3,1)$   rotation in
 the tangent space  and thus it imparts no constraint on the gauge degree of freedom of the
   Lagrangian.
 %%%%%%%%%%%%%%%%%%%%%%%%%%%%%%%%%%%%%%%%%%%%%%%%%%%%%%%%%%%%%%%%%%%%%%%%%%%%%%%%%%%
 \section{Scalar Field and Spinorial Matter}

 Now we are in a position to write the total gravity Lagrangian in the presence of a spinorial matter field, given
by\begin{eqnarray}
\mbox{$\mathcal{L}$}_{tot.}&=&\mbox{$\mathcal{L}$}_{G}+\mbox{$\mathcal{L}$}_{D}, \label{eqn:apr}
\end{eqnarray} where
\begin{eqnarray}
\mbox{$\mathcal{L}$}_{D}&=&\phi^2[\frac{i}{2}\{\overline{\psi}{}^*\gamma\wedge D\psi+\overline{D\psi}\wedge{}^*\gamma\psi\}-\frac{g}{4}\overline{\psi}\gamma_5\gamma\psi\wedge T\nonumber\\&{}&+c_\psi\sqrt{{}^*dT} \overline{\psi}\psi\eta]\label{eqn:apr1}    \\
    \gamma_\mu &:=&\gamma_a e^a{}_\mu,\hspace{2mm}
    {}^*\gamma:=\gamma^a\eta_a,\hspace{2mm}
    D:=d+\Gamma\\
    \Gamma &:=&\frac{1}{4}\gamma^\mu D^{\{\}}\gamma_{\mu}=\frac{1}{4}\gamma^\mu \gamma_{\mu:\nu}dx^\nu\nonumber\\&=&-\frac{i}{4}\sigma_{ab}e^{a\mu}e^b{}_{\mu:\nu}dx^\nu
\end{eqnarray}
here $D^{\{\}}$, or $:$ in tensorial notation, is Riemannian torsion free covariant differentiation acting on external indices only; $\sigma^{ab}=\frac{i}{2}(\gamma^a\gamma^b-\gamma^b\gamma^a)$,  $\overline{\psi}=\psi^\dag\gamma^0$ and $g$, $c_\psi$ are both dimensionless coupling constants. Here $\psi$ and $\overline{\psi}$ have dimension $(length)^{-\frac{1}{2}}$ and conformal weight $-\frac{1}{2}$. It can be verified that under $SL(2,C)$ transformation on the spinor field and gamma matrices, given by,
\begin{eqnarray}    \psi\rightarrow\psi^\prime&=&S\psi,\hspace{2mm}\overline{\psi}\rightarrow\overline{\psi^\prime}=\overline{\psi}S^{-1}\nonumber\\\mbox{and}\hspace{2mm}\gamma\rightarrow\gamma^\prime&=&S\gamma S^{-1},
\end{eqnarray}where \mbox{$S=\exp(\frac{i}{4}\theta_{ab}\sigma^{ab})$,}
$\Gamma$ obeys the transformation property of a $SL(2,C)$ gauge connection, i.e.
\begin{eqnarray}
    \Gamma\rightarrow\Gamma^\prime&=&S(d+\Gamma)S^{-1}\\
    \mbox{s. t.}\hspace{2mm}D\gamma&:=&d\gamma+[\Gamma,\gamma]=0.\label{eqn:a47}
\end{eqnarray}Hence   $\gamma$ is a covariantly constant matrix valued one form w. r. t. the \mbox{$SL(2,C)$} covariant derivative $D$. By Geroch's theorem\cite{Ger68} we know that - the existence of  the spinor structure is equivalent to the existence of a global field of orthonormal tetrads on the space and time orientable manifold. Hence use of $\Gamma$ in the $SL(2,C)$ gauge covariant derivative is enough in a Lorentz invariant theory where de Sitter symmetry is broken.

In  appendices A and  B of Ref.\cite{Mah07a},  by varying  the independent fields in the Lagrangian $\mbox{$\mathcal{L}$}_{tot.}$, we obtain the Euler-Lagrange equations and then after some simplification we get the following results of this section
\begin{eqnarray}
\bar{\nabla} e_{a}=0,\label{eqn:abc9}\\
{}^*N=\frac{6}{\kappa},\label{eqn:abc17}
\end{eqnarray}
 i.e. $\bar{\nabla}$ is torsion free and $\kappa$ is an integration constant having  dimension of $(length)^{2}$.\footnote{In (\ref{eqn:abcd1}), $\bar{\nabla}$ represents a $SO(3,1)$ covariant derivative, it is only on-shell torsion-free through the field equation (\ref{eqn:abc9}). The $SL(2,C)$ covariant derivative represented by the operator $D$ is torsion-free by definition, i.e. it is torsion-free both on on-shell and off-shell. Simultaneous and independent use of both $\bar{\nabla}$ and $D$ in the Lagrangian density (\ref{eqn:apr}) has been found to be advantageous in the approach of this article. This amounts to the emergence of the gravitational constant   $\kappa$ to be  only an on-shell  constant and this justifies the need for the introduction of the Lagrangian multiplier $b_a$ which appears twice in the Lagrangian density (\ref{eqn:abcd1}) such that $\bar{\omega}^a{}_b$ and $e^a$  become independent fields.}
\begin{eqnarray}    m_\psi=c_\psi\sqrt{{}^*dT}=\frac{c_\psi}{\sqrt{\kappa}},\label{eqn:a59}
\\i{}^*\gamma\wedge D\Psi-\frac{g}{4}\gamma_5\gamma\wedge T\Psi+m_\Psi\Psi\eta=0,\nonumber\\
i\overline{D\Psi}\wedge{}^*\gamma-\frac{g}{4}\overline{\Psi}\gamma_5\gamma\wedge T+m_\Psi\overline{\Psi}\eta=0,  \label{eqn:a580}
\end{eqnarray}  where $\Psi=\phi\psi$ and $ m_\Psi= m_\psi$.

\begin{eqnarray} G^b{}_a\eta&=&-\kappa[ \frac{i}{8}\{\overline{\Psi}(\gamma^b D_a+\gamma_a D^b)\Psi-(\overline{D_a\Psi}\gamma^b+\nonumber\\&{}&\overline{D^b\Psi}\gamma_a)\Psi\}\eta-\frac{g}{16}\overline{\Psi}\gamma_5(\gamma_a {}^*T^b+\gamma^b {}^*T_a)\Psi\eta\nonumber\\&{}&+f\partial_a\phi\partial^b\phi\eta+\frac{1}{2}(h)\eta\delta^b{}_a],\label{eqn:e1}\\0&=&[\frac{1}{2}  \overline{\nabla}_\nu\overline{\Psi}\{\frac{\sigma^b{}_{a}}{2}, \gamma^\nu\} \Psi+\frac{i}{2}\{\overline{\Psi}(\gamma^b D_a-\gamma_a D^b)\Psi\nonumber\\&{}&-(\overline{D_a\Psi}\gamma^b-\overline{D^b\Psi}\gamma_a)\Psi\} \nonumber\\&{}&-\frac{g}{4}\overline{\Psi}\gamma_5(\gamma_a {}^*T^b-\gamma^b {}^*T_a)\Psi]\eta,\label{eqn:e2}
\end{eqnarray}

\begin{eqnarray} \kappa d[\frac{g}{4}{}^*(\overline{\Psi}\gamma_5\gamma\Psi\wedge T)-f{}^*(d\phi\wedge{}^*d\phi)+2h-\frac{\beta}{\kappa}\phi^2] \nonumber\\{}=-\frac{g}{4}\overline{\Psi}\gamma_5\gamma\Psi,\label{eqn:abc141}
    \\ \frac{2}{\kappa}\beta \phi + f^\prime(\phi)d\phi\wedge{}^*d\phi -h^\prime(\phi)
\eta+2fd{}^*d\phi\nonumber\\=-2\phi[\frac{i}{2}\{\overline{\psi}{}^*\gamma\wedge D\psi+\overline{D\psi}\wedge{}^*\gamma\psi\}\nonumber\\\hspace{4mm}-\frac{g}{4}\overline{\psi}\gamma_5\gamma\psi\wedge T+m_\psi \overline{\psi}\psi\eta]=0.\label{eqn:abc142}
\end{eqnarray}

 In Ref.\cite{Mah07a} the following comments were made,
%\begin{enumerate}
\begin{itemize}
\item Right hand side of equation (\ref{eqn:e1}) may be interpreted\cite{Bor02} as  ($-\kappa$) times the energy-momentum stress tensor of the Dirac field $\Psi(\overline{\Psi})$ together with the scalar field $\phi$. Where by equation
(\ref{eqn:abc17}) the gravitational constant $\kappa$ is
$\frac{6}{{}^*N}$ and then by equation (\ref{eqn:a59}) mass of the
spinor field is proportional to $\sqrt{{}^*N}$.
\item Equation (\ref{eqn:e2})  represents covariant conservation of angular momentum of the Dirac field in the Einstein-Cartan space $U_4$ as a generalization of the same in the Minkwoski space $M_4$\cite{Itz85}.
\item Equation (\ref{eqn:abc142}) is the field equation of the scalar field $\phi$. Here it appears that, in the on-shell, other than gravity, it has no source. Whereas in equation (\ref{eqn:abc141}), there is a non trivial appearance of the torsion, the axial-vector matter-current and the scalar field $\phi$; provided the coupling constant $g$ is not negligible in a certain energy scale.
%\end{enumerate}
\end{itemize}

 In Ref.\cite{Mah07a}, the spin-torsion interaction term was neglected, although it might be that  $g$ played a dominant role in the early universe. In other words, we may say that the scalar field, which appears to be connected with the spinor field only in equation (\ref{eqn:abc141}), is at present playing the role of the dark matter and/or dark radiation.   The consequence of this  spin-torsion interaction term, in the very early universe,   may be linked to the cosmological inflation without false vacuum\cite{Gas86}, primordial density fluctuation \cite{And99a,Pal99} and/or to the repulsive gravity\cite{Gas98}.
%%%%%%%%%%%%%%%%%%%%%%%%%%%%%%%%%%%%%%%%%%%%%%%%%%%%%%%%%%%%%%%%%%%%%%%%%%%%%%%%%%%
 \section{($g=0$) $\Rightarrow$ Standard FRW Cosmology with Dark Matter}
From Ref.\cite{Mah07a} we may repeat, in short, the analysis of the results obtained in section-3. In the background of a FRW-cosmology
where the metric tensor is given by
\begin{eqnarray}
 g_{00}=-1 ,\hspace{2mm}g_{ij}= \delta_{ij}a^2(t)  \hspace{2mm}\mbox{where}\hspace{2mm}
i,j=1,2,3;\label{eqn:aay}
\end{eqnarray}such that
\begin{eqnarray}
    &{}&e=\sqrt{-\det(g_{\mu\nu})}= a^3\label{eqn:zaz}
\end{eqnarray}
Taking $g=0$, we get Einstein's equations of standard FRW Cosmology, where non-vanishing components of Einstein's tensor (\ref{eqn:e1}), w. r. t. external indices,
are given by
\begin{eqnarray}
    G^{0}{}_{0}&=&-3(\frac{\dot{a}}{a})^2=-\kappa(\rho_{BM}+\frac{\beta}{\kappa}\phi^{2}+\frac{\lambda}
{\kappa}-\frac{3h}{2}) \nonumber\\G^{j}{}_{i}&=&-(\frac{2\ddot{a}}{a}+\frac{\dot{a}^2}{a^2})
\delta^{j}{}_{i}=-\kappa\frac{1}{2}(h)\delta^{j}{}_{i}\label{eqn:11a}
\end{eqnarray}where we have assumed that, in the cosmic scale, the observed (luminous) mass distribution is baryonic and co moving, s. t.
\begin{eqnarray}
    &{}&\mathop\sum\limits_{\Psi}^{}\frac{i}{8}\{\overline{\Psi}(\gamma^b D_a+\gamma_a D^b)\Psi-(\overline{D_a\Psi}\gamma^b+\overline{D^b\Psi}\gamma_a)\Psi\}\nonumber\\&{}&\hspace{6mm}=\rho_{BM}=\frac{M_{BM}}{V},\hspace{2mm}\mbox{for $a=b=0$},\nonumber\\&{}&\hspace{6mm}=0,\hspace{2mm}\mbox{otherwise}.
\end{eqnarray}Here $M_{BM}$ and $V$ are the total baryonic mass and volume of the universe respectively.

From the forms of $G^{0}{}_{0}$ and $G^{j}{}_{i}$ it appears that the term $\frac{\beta}
{\kappa}\phi^{2}$ represents pressure-less energy density i.e. $\phi^{2}\propto a^{-3}\propto
 \frac{1}{e}$. Following Ref.\cite{Mah07a} equation   (\ref{eqn:11a}) reduces to
\begin{eqnarray}
G^{0}{}_{0}&=&-3(\frac{\dot{a}}{a})^2=-\kappa(\rho_{BM}+\rho_{DM}+\rho_{DR}+\rho_{VAC.}) \nonumber\\G^{j}{}_{i}&=&-(\frac{2\ddot{a}}{a}+\frac{\dot{a}^2}{a^2})
\delta^{j}{}_{i}\nonumber\\&=&\kappa(p_{BM}+p_{DM}+p_{DR}+p_{VAC.}) \label{eqn:11b}\delta^{j}{}_{i},
\end{eqnarray}  where
\begin{eqnarray}
    p_{BM}=p_{DM}=0\\
    \rho_{DM}=\frac{\beta}{\kappa}\phi^2\\
    \rho_{DR}=\frac{3\gamma}{2}\phi^{\frac{8}{3}},\hspace{2mm}p_{DR}=\frac{1}{3}\rho_R\\
\rho_{VAC.}=-p_{VAC.}=\frac{\lambda}{4\kappa}=\Lambda \mbox{ (say)}.
\end{eqnarray}As the scalar field $\phi$, at present scale, appears to be non-interacting with the spinor field $\Psi$, \textit{vide equations} (\ref{eqn:a580}), (\ref{eqn:abc141}) \& (\ref{eqn:abc142}), the quantities having subscripts ${}_{BM}$, ${}_{DM}$,   ${}_{DR}$ and ${}_{VAC.}$ may be assigned to the baryonic matter, the dark matter, the dark radiation and the vacuum energy respectively. If we add another Lagrangian density to (\ref{eqn:apr1}) corresponding to the Electro-Magnetic field and modify $D$ by $D+A$, where $A$ is the $U(1)$   connection one form, and also consider massless spinors having $c_\psi=0$, then on the r. h. s. of equations in (\ref{eqn:11b}),  $\rho_{DR}$ and $p_{DR}$ would be replaced by $\rho_R$ and $p_{R}$  containing  various  radiation components, given by
\begin{eqnarray}
    \rho_R&=&\rho_{DR}+\rho_\gamma+\rho_\nu ,\nonumber\\
    p_R&=&p_{DR}+p_\gamma+p_\nu\\
    \mbox{s. t.}\hspace{2mm}p_R&=&\frac{1}{3}\rho_R ,
\end{eqnarray}where the subscripts have their usual meanings.
%%%%%%%%%%%%%%%%%%%%%%%%%%%%%%%%%%%%%%%%%%%%%%%%%%%%%%%%%%%%%%%%%%
\section{($g\neq0$) $\Rightarrow$ ??}

Here we consider the case where  $g$ is not negligible in equation (\ref{eqn:abc141}).
From this equation    we can define 
 \begin{eqnarray} J^A&\equiv&\overline{\Psi}\gamma_5{}^*\gamma\Psi\nonumber\\&=&- \kappa {}^*d[{}^*(\overline{\Psi}\gamma_5\gamma\Psi\wedge T)-\frac{4f}{g}{}^*(d\phi\wedge{}^*d\phi)+\frac{8}{g}h-\frac{4\beta}{g\kappa}\phi^2],\nonumber\\&\equiv & {}^*d\chi \hspace{2mm}\mbox{(say)}\label{eqn:abc145}
\end{eqnarray} Then incorporating quantum mechanical corrections we get the  axial anomaly \cite{Cha97,Fuj04,Iof08}
\begin{eqnarray}
\square\chi\equiv d{}^*d\chi=dJ^A	=&{}& 2im{}^*(\overline{\Psi}\gamma_5\Psi)+\frac{1}{8\pi^2}[R^{ab}\wedge R_{ab}+\nonumber\\&{}& \underbrace{F\wedge F}_{\mbox{\scriptsize{abelian}}}+\mbox{Tr}({\underbrace{G\wedge G}_{\mbox{\scriptsize{non-abelian}}}})]+\textbf{T}\equiv\rho\hspace{2mm}\mbox{(say)}\label{eqn:abc146}
\end{eqnarray}where $\textbf{T}$ is torsional contribution to the anomaly\cite{Cha97} and $F$ \& $G$ are the abelian  \& the non-abelian gauge field strengths respectively.
This equation may be considered to be the four dimensional generalization of Poisson equation where the density four form $\rho$ acts as the source of the scalar field $\chi$.
\subsection{Isotropy and Homogeneity of Standard FRW Cosmology}

In previous section we see that neglect of $g$ leads to the standard FRW cosmology. Therefore if we assume,  without $g$ being negligible, that the back ground geometry may be extrapolated from that of  standard FRW geometry   then taking clue from equation (\ref{eqn:abc141}), we may 
postulate
\begin{eqnarray}f{}^*(d\phi\wedge{}^*d\phi)-2h+\frac{\beta}{\kappa}\phi^2=constant.\label{eqn:abc143}
\end{eqnarray}
In this case equation (\ref{eqn:abc141}) reduces to
\begin{eqnarray}
	\overline{\Psi}\gamma_5\gamma\Psi=- \kappa d{}^*(\overline{\Psi}\gamma_5\gamma\Psi\wedge T).
\end{eqnarray}
Now defining an axial vector current three-form, given by
\begin{eqnarray}
	J^A_1 = 	
\kappa{}^*(\overline{\Psi}\gamma_5\gamma\Psi\wedge T)T,\label{eqn:abc144}
	\end{eqnarray}	and using $(\ref{eqn:abc17})$ we get
	\begin{eqnarray}
	dJ^A_1 = 0
\end{eqnarray}
\subsection{General case}
In this case, with out assuming any particular background geometry, we may define another axial vector current three-form, given by
\begin{eqnarray}
	J^A_2 = 	
\kappa(\overline{\Psi}\gamma_5\gamma\Psi\wedge F)\label{eqn:abc147}
	\end{eqnarray}
	where $F=dA$ is the electro-magnetic or any $U(1)$ field strength. Then using equation (\ref{eqn:abc141}) we get 	
\begin{eqnarray}
	dJ^A_2 = 0
\end{eqnarray}
%%%%%%%%%%%%%%%%%%%%%%%%%%%%%%%%%%%%%%%%%%%%%%%%%%%%%%%%%%%%%%%%%%
\section{Discussion}

Here we see that if we introduce a scalar field $\phi$ to cause  the de Sitter connection to have the proper dimension of a gauge field and also link this scalar field with the dimension of a Dirac field then we find that the Euler-Lagrange equations of both the fields to be mutually non-interacting. But they are indirectly connected to each other when we consider Euler-Lagrange equations of other geometric fields such as torsion and tetrad.  Variation of the $SO(3,1)$ spin connection as an entity independent of the tetrads we get the Newton's constant as inversely proportional to the topological Nieh-Yan density  and then the mass of the spinor field has been shown to be linked to the Newton's constant. Then using symmetries of the Einstein's tensor we get covariant conservation of angular momentum of the Dirac field in the particular class of geometry in  $U_4$ as a generalization of the same in the Minkwoski space $M_4$. Neglecting the  spin-torsion interaction term and  considering FRW cosmology we are able to derive standard cosmology with standard energy density together with dark matter, dark radiation and cosmological constant.

Here we see that variation of torsion in the action gives us the axial vector one-form $j_5=\overline{\Psi}\gamma_5\gamma\Psi$ to be an exact form [Eqn. (\ref{eqn:abc141})]. Divergence of the matter axial current, the dual of $j_5$, gives us the four dimensional generalization of Poisson equation where the density four form $\rho$ acts as the source of the scalar field $\chi$ [Eqn. (\ref{eqn:abc146})]. If we consider FRW geometry to be in the background then the FRW postulate [Eqn. (\ref{eqn:abc143})] makes it possible to define an axial vector current three-form $	J^A_1$ as a product of torsion and matter current $j_5$. $	J^A_1$ is conserved and as well as gauge invariant. In  manifolds having arbitrary background geometry the product of $j_5$ and an $U(1)$ field strength $F$ gives us another gauge invariant conserved current $	J^A_2$. These conserved axial currents implies pseudoscalar conserved charges. Being not Noether charges, physical significances of these charges are being studied and will be reported in a forthcoming paper.

 \section*{Acknowledgment}
           I wish to thank Prof. Pratul Bandyopadhyay, Indian Statistical Institute,  Kolkata,  for his valuable remarks and fruitful suggestions on this
           problem.

%%%%%%%%%%%%%%%%%%%%%%%%%%%%%%%%%%%%%%%%%%%%%%%%%%%%%%%%%%%%%%%%%%%%%%%%%%%%%%%%%%%%%%%%%%%%%%%%%%%%%%%%%%%%%%%%%%%%%%%%
\renewcommand{\theequation}{\mbox{A}\arabic{equation}}
\setcounter{equation}{0}
%\end{thebibliography}
%\bibliographystyle{unsrt}
%\bibliography{tbib}

\begin{thebibliography}{10}

\bibitem{Mah02a}
P.~Mahato.
\newblock {\em Mod. Phys. Lett.}, A17:1991, 2002.

\bibitem{Con94}
A.~Connes.
\newblock {\em Noncommutative Geometry}.
\newblock Academic Press, New York, 1994.

\bibitem{Gho00}
P.~Ghosh and P.~Bandyopadhyay.
\newblock {\em Int. J. Mod. Phys.}, A15:3287, 2000.

\bibitem{Mah04}
P.~Mahato.
\newblock {\em Phys. Rev.}, D70:124024, 2004.

\bibitem{Mah05}
P.~Mahato.
\newblock {\em Int. J. Theor. Phys.}, 44:79, {2005 (arXiv:gr-qc/0603109)}.

\bibitem{Mah07a}
P.~Mahato.
\newblock {\em Int. J. Mod. Phys.}, A 22:835, {2007 (arXiv:gr-qc/0603134)}.

\bibitem{Mah07b}
P.~Mahato.
\newblock {\em Ann. Fond. Louis de Broglie}, 32:297, 2007.

\bibitem{Lob07a}
{C. G. B\"ohmer, L. Hollenstein and F. S. N. Lobo}.
\newblock {\em Phys. Rev.}, D76:084005, 2007.

\bibitem{Fuj04}
{K. Fujikawa and H. Suzuki}.
\newblock {\em Path Integrals and Quantum Anomalies}.
\newblock Clarendon Press, Oxford, 2004.

\bibitem{Egu80}
{T. Eguchi, P. B. Gilkey} and A.~J. Hanson.
\newblock {\em Phys. Rep.}, 66:213, 1980.

\bibitem{Kim69}
T.~Kimura.
\newblock {\em Prog. Theo. Phys.}, 42:1191, 1969.

\bibitem{Sal72}
{R. Delbourgo and A. Salam}.
\newblock {\em Phys. Lett.}, 40B:381, 1972.

\bibitem{Fre76}
{T. Eguchi and P. Freund}.
\newblock {\em Phys. Rev. Lett.}, 37:1251, 1976.

\bibitem{Wit84}
{L. Alvarez-Gaum´e and E. Witten}.
\newblock {\em Nucl.Phys.}, B234:269, 1984.

\bibitem{Ton86}
{L. Bonora, P. Pasti and M. Tonin}.
\newblock {\em J. Math. Phys.}, 27:2259, 1986.

\bibitem{Zan91}
{A. Mardones and J. Zanelli}.
\newblock {\em Class. Quantum Grav.}, 8:1545, 1991.

\bibitem{Cha97}
O.~Chandia and J.~Zanelli.
\newblock {\em Phys. Rev.}, D55:7580, 1997.

\bibitem{Wei96}
S.~Weinberg.
\newblock {\em The Quantum Theory of Fields, Vol. II}.
\newblock Cambridge University Press, Cambridge, 1996.

\bibitem{Kak93}
M.~Kaku.
\newblock {\em Quantum Field Theory}.
\newblock Oxford University Press, Oxford, 1993.

\bibitem{Car22}
E.~Cartan.
\newblock {\em Ann. Ec. Norm.}, 40:325, 1922.

\bibitem{Car24}
E.~Cartan.
\newblock {\em Ann. Ec. Norm.}, 1:325, 1924.

\bibitem{Che74}
S.~Chern and J.~Simons.
\newblock {\em Ann. Math.}, 99:48, 1974.

\bibitem{Che71}
S.~Chern and J.~Simons.
\newblock {\em Proc. Natl. Acad. Sci. (USA)}, 68:791, 1971.

\bibitem{Nie82}
H.~T. Nieh and M.~L. Yan.
\newblock {\em J. Math. Phys.}, 23:373, 1982.

\bibitem{Ger68}
R.~Geroch.
\newblock {\em J. Math. Phys.}, 9:1739, 1968.

\bibitem{Bor02}
V.~Borokhov.
\newblock {\em Phys. Rev.}, D65:125022, 2002.

\bibitem{Itz85}
C.~Itzykson and J.~Zuber.
\newblock {\em Quantum Field Theory}.
\newblock McGraw Hill Book Company, Singapore, 1985.

\bibitem{Gas86}
M.~Gasperini.
\newblock {\em Phys. Rev. Lett.}, 56:2873, 1986.

\bibitem{And99a}
L.~C.~Garcia de~Andrade.
\newblock {\em Phys. Lett.}, B468:28, 1999.

\bibitem{Pal99}
D.~Palle.
\newblock {\em Nuovo cim.}, B114:853, 1999.

\bibitem{Gas98}
M.~Gasperini.
\newblock {\em Gen. Rel. Grav.}, 30:1703, 1998.

\bibitem{Iof08}
B.~L. Ioffe.
\newblock {\bf Axial anomaly in quantum electro- and chromodynamics and the
  structure of the vacuum in quantum chromodynamics}.
\newblock {\em Report at the Session of the Physical Division of Russian
  Academy of Sciences, devoted to 100 years L.D.Landau birthday, 22-23 January
  2008,}, arXiv:0809.0212v1[hep-ph], 2008.

\end{thebibliography}
                
%\end{document}               

\end{document}